# Co-operation of two-state processes and 1/f noise

Giovanni Zanella

*Dipartimento di Fisica dell'Università di Padova and Istituto Nazionale di Fisica Nucleare, Sezione di Padova, via Marzolo 8, 35131 Padova, Italy*

A general physical model is presented for *1/f* noise. The main questions raised by this type of noise can be solved if at the origin of the phenomenon we consider many similar like RTN two-state processes in co-operation among them to generate a Gaussian average process with *1/f* power spectrum.

Even if the originating RTN processes have the same relaxation time, their co-operation produces in short times secondary RTN processes with a distribution of relaxation times which generates again *1/f* noise.

An extension of the model to a single two-state process viewed in "series" reveals once more the appearance of *1/f* noise.

Experimental results, found in the literature, agree with this model under various aspects.

## 1. Introduction

Many natural processes appear with fluctuations whose the measured power spectrum *S(f)* increases with decreasing frequency *f* , approximately as $1/f^g$ where $\boldsymbol{g}$ is a positive number close enough to one. These phenomena are related to the so-called *1/f* noise also known as "flicker" noise or "pink" noise .

Despite great progress being made in *1/f* noise physics the source of these fluctuations remain unknown in the case of most systems and the problem remains largely unsolved in its generality.

It would be too lengthy to report all references concerning *1/f* noise but an extensive bibliography has been drawn up in [1].

The principal feature of *1/f* noise is that *S(f)* increases with decreasing frequency *f* down to the lowest possible frequencies for conducting measurements. This noise is therefore spectrally *scaling* that is it is statistically identical to its transformation by contraction in time, or another independent variable, followed by a corresponding change in intensity. This scaling property is typical of the *fractals* [2].

The aim of this contribution is to put forward an explanation to the problem based on a general physical mechanism which reveals the origin and essence of the effect.

It is common knowledge that the "intense" events are rare because they are due to the "coincidence" (or co-operation) of various favourable (or adverse) circumstances, so that the probability of occurrence, and then the frequency, decreases inversely with the power.

Thanks to the existence, in nature, of microscopic or macroscopic two-state processes like random telegraph noise (RTN), we can see, on the basis of the Gaussian statistics intended as a limit of the binomial statistics, that *1/f* noise can be generated. This is due to the coincidences of many similar RTN processes viewed "in parallel" or by a single binary process treated "in series" on the basis of the same statistics.

In the following, we refer only to time phenomena.

## 2. Co-operation of RTN processes and 1/f noise

The idea that *1/f* noise can be generated by a superposition of several RTN statistically independent processes was suggested by M. Sourdin to explain the "flicker effect" in a thermoionic current [3]. We recall this model here [1].

In a RTN process, where the variable $u$ ranges between the values $u_1$ and $u_2$, the power spectrum $S_u(w)$ is a Lorentzian:

$$S_u(w) = (u_1 - u_2)^2 \, p_1 p_2 \, \frac{4t}{1 + w^2 t^2} \qquad (1)$$

where $p_1$ is the probability of finding the value $u_1$, $p_2 = 1 - p_1$ is the probability of finding the value $u_2$ and $t$ is the "relaxation time" (the average time spent in each state before making a transition to another state, or the inverse of the total rate of transitions, back and forth, in the process).

The term $(u_1 - u_2)^2 \, p_1 p_2$ of equation (1) represents the variance of the process [1].

In the case of a superposition of many RTN processes with a continuous distribution of relaxation times, $S_u(w)$ has to be averaged over this distribution with a weight function $p(t)dt$ which includes the contribution to the variance of processes whose relaxation times lie in the interval from $t$ to $t + dt$. In practice

$$S_u(w) = \int_0^\infty p(t) \frac{4t}{1 + w^2 t^2} dt \quad . \qquad (2)$$

Supposing we have $p(t) \propto 1/t$ in some interval from $t_{min}$ to $t_{max} \gg t_{min}$ and zero outside this interval, one has $S_u(w) \propto 1/w$ in the frequency range $t_{max}^{-1} \ll w \ll t_{min}^{-1}$ [1].

Instead, if $w \ll t_{max}^{-1}$, $S_u(w) \approx const.$ [1][3].

As this model requires a continuous distribution of relaxation times on a wide range, it is hard to explain physically the origin of such distribution for a large diversity of systems.

A different approach to the problem is possible considering the "coincidences" which appear in the superposition of many RTN processes with about the same probability of occurrence and the same relaxation time $t_{min}$. These coincidences correspond to the number of contemporary high (or low) levels which appear during a time $t_{min}$.

The statistics which governs the probability $P(c)$ of the number of coincidences is the Gaussian distribution of probability, for we are in a situation of binomial distribution with a large number of attempts and a probability $p$ for the events which is not very small. In our case the number of attempts corresponds to the number of involved primary RTN processes.

Consequently we can write:

$$P(c) = \frac{1}{s_c \sqrt{2p}} e^{-\frac{(c-<c>)^2}{2s_c^2}} \quad , \qquad (3)$$

where $<c>$ denotes the mean of $c$ and $s_c^2$ represents the variance of the $P(c)$ distribution.

If we introduce the fluctuation $\Delta c = c - <c>$ from the mean, equation (3) can be simply rewritten as:

$$P(\Delta c) \propto e^{-\frac{\Delta c^2}{2s_{\Delta c}^2}} \quad , \qquad (4)$$

where $P(\Delta c) = P(c)$ and $s_{\Delta c}^2 = s_c^2$.

Each of the concurrent processes can induce linear variations on a physical variable $u$ so $\Delta u$ is the fluctuation of $u$ from $\langle u \rangle$ proportional to $\Delta c$.

The fluctuation $\Delta u$ will also follow Gaussian distribution:

$$P(\Delta u) \propto e^{-\frac{\Delta u^2}{2s_{\Delta u}^2}} \quad , \qquad (5)$$

where $s_{\Delta u}^2$ denotes the variance of the $P(\Delta u)$ distribution.

We can solve the equation (5) in the unknown $\Delta u^2$ and obtain

$$\Delta u^2 = -2 s_{\Delta u}^2 (\ln P(\Delta u) + const) \quad . \qquad (6)$$

With reference to equation (6) $P(\Delta u) = P(\Delta u)_{max}$ when $\Delta u^2 = 0$ so $const = - \ln P(\Delta u)_{max}$ and

$$\Delta u^2 = 2 s_{\Delta u}^2 \ln \frac{P(\Delta u)_{max}}{P(\Delta u)} \quad . \qquad (7)$$

$\Delta u^2$ of equation (7) represents the average power of the Gaussian process concerning the probability $P(\Delta u)$. This average power seen in the time-domain is due to the concourse of a precise number of originating RTN processes which coincide in time $t_{min}$ and generate fluctuation $\Delta u$.

The *law of large numbers* allows us to introduce an average frequency *f* of the outcomes, proportional to their probabilities of occurrence, thereby enabling us to substitute $P(\Delta u)$ with *f* in equation (7). We can conceive

$$f = \frac{P(\Delta u)}{t_{min}} \quad \text{and} \quad f_{max} = \frac{P(\Delta u)_{max}}{t_{min}}. \tag{8}$$

Therefore

$$\Delta u^2 = 2\sigma_{\Delta u}{}^2 \ln \frac{f_{max}}{f} \quad . \tag{9}$$

It is important to note that in this context we tacitly admit that $t_{min}$ also represents the maximum time employed for one measurement. In this sense the process is analysed in its intrinsic time-scale.

The power $\Delta u^2$ of equation (9) represents the maximum power of the deterministic process of frequency *f* to which the Gaussian probabilistic process converges when we consider its average behaviour. On the other hand $\Delta u^2$ is also an average power in the time interval $t_{min}$.

Various components of different frequency will generate the power $\Delta u^2$. Components with a frequency lower than *f* are not possible because this process is periodical and it can be expanded by a Fourier series whose fundamental frequency is just *f*. The maximum frequency of these components will be the maximum frequency of the process, that is the $f_{max}$ of the relationship (8).

Subsequently, in the hypothesis of a large number of RTN "fluctuators" and introducing the power spectrum $S_u(f)$, we have:

$$\int_f^{f_{max}} S_u(f)\,df = \Delta u^2 \quad . \tag{10}$$

If $S_u(f) = 2\sigma_u{}^2/f$ we obtain:

$$\int_f^{f_{max}} \frac{2\sigma_u{}^2}{f}\,df = 2\sigma_u{}^2 \ln \frac{f_{max}}{f} = \Delta u^2 \quad , \tag{11}$$

so

$$S_u(f) \propto \sigma_u{}^2/f . \tag{12}$$

The Gaussian nature of *1/f* noise is proved in most experimental observations on its statistical properties, and particularly by experiments performed by R.F. Voss on different solid state devices [4]. In particular R.F. Voss tested the correlation between Gaussian behaviour and *1/f* noise observed in a sufficiently pure form.

## 3. Filtering effects

If the time employed for one measurement of $\boldsymbol{D}u$ is $\boldsymbol{D}t > \boldsymbol{t}_{min}$ and $\boldsymbol{D}u$ is sampled with frequency *1/Dt*, thus the generating processes do not appear as pure RTNs. In any case, these processes have the same probability distribution so their summation always has a Gaussian distribution thanks to the *central limit theorem*.

Applying the *law of large numbers* the frequency of the fluctuation $\boldsymbol{D}u$ will be:

$$f^* = \frac{P(\Delta u)}{\Delta t} \qquad \text{and} \qquad f^*_{max} = \frac{P(\Delta u)_{max}}{\Delta t} \quad . \tag{13}$$

Therefore, likewise in the equation (9),

$$\Delta u^2 = 2\boldsymbol{s}_{\Delta u}^2 \ln \frac{f^*_{max}}{f^*} \quad . \tag{14}$$

$\boldsymbol{D}u^2$ of equation (14) denotes the maximum power of the deterministic process of frequency $f^*$ to which the Gaussian probabilistic process converges when we consider its average behaviour for a long time.

Now, $\boldsymbol{D}u^2$ represents an average power in the time interval $\boldsymbol{D}t$ and will be generated in the frequency-domain by spectral components with a frequency which ranges from the minimum $f^*$ ($\boldsymbol{D}u$ is related to a periodical process of frequency $f^*$) to the maximum $f_{max}$ as reported in the equation (8).

Thus, in the hypothesis of a large number of "fluctuators" and introducing the power spectrum $S_u(f)$ we have:

$$\int_{f^*}^{f_{max}} S_u(f)\, df = \Delta u^2 = 2\boldsymbol{s}_{\Delta u}^2 \ln \frac{f^*_{max}}{f^*} \quad . \tag{15}$$

Therefore, being as $f^*_{max}$ is generally different from $f_{max}$, $S_u(f)$ cannot be a pure *1/f* process as in the case of the generating RTN processes.

From another point of view, the sampling of the variable $\boldsymbol{D}u$ by a time interval $\boldsymbol{D}t$ (repeated with a frequency *1/Dt*) filters the power spectrum *1/f* by a *sinc*$^2$ operator [7], that is

$$S_u(f) \propto \frac{1}{f} \frac{sin^2(\boldsymbol{p}f\Delta t)}{(\boldsymbol{p}f\Delta t)^2} \quad . \tag{16}$$

So, $S_u(f)$ becomes $\boldsymbol{\mu}$ *1/f* when $f \boldsymbol{D}t \circledR 0$.

## 4. Distribution of relaxation times and 1/f noise

The *1/f* process can be decomposed for short times even in secondary RTN processes. These secondary processes will have the amplitude $du = Du_i - Du_{i-1} = const$ for every $i$ , being as $Du_i$ is the fluctuation of $u$ corresponding to the fluctuation $Dc_i$ . So, to each fluctuation $Dc_i$ will correspond one secondary RTN process whose relaxation time $t$ will be related to the average frequency of $Dc_i$ (in practice $t = 1/2f$). The power spectrum of these secondary RTN processes will be:

$$S_{u,t}(w) = du^2 \ p_1(t) \ p_2(t) \frac{4t}{1 + w^2 t^2} \qquad (17)$$

where $du^2 p_1(t) p_2(t)$ is the variance of the process.

Necessarily, $S_u(w) = S_t \ S_{u,t}(w)$ and we can demonstrate $S_t \ S_{u,t}(w) \ \mu \ 1/w$ by operating in the same manner as paragraph 2.

In fact, in the case of a continuous distribution of the relaxation times $S_{u,t}(w)$ has to be averaged by a weight function $p(t)dt$ which includes the contribution to the variance of processes whose relaxation times lie in the interval from $t$ to $t + dt$.

We can suppose $p(t) \mu 1/t$, thereby obtaining the power spectrum of the resulting process by integrating $S_{u,t}(w)$ within the interval $t_{min}, t$ , that is

$$S_u(w) \ \mu \ 1/w \qquad , \qquad (18)$$

if $t^{-1} << w << t_{min}^{-1}$ .

In conclusion, the distribution of the RTN processes which determines the *1/f* noise is not only a mathematical point of curiosity, but it is precisely the physical consequence of the interaction of similar originating RTN processes which forces the process to have a *1/f* power spectrum, even for short times.

## 5. Dependence of 1/f noise on mean voltage or current

As we know, in uniform conductors the voltage power spectrum of *1/f* noise is proportional to the square of the steady current flowing through the sample or the mean voltage $<V>$ across the same sample in the region in which Ohm's law is obeyed.

These relations have been verified many times on various conductors. For instance, they were verified for continuous metal films by Voss & Clarke [5]. These authors have also proved [6] that the resistance $R$ fluctuates under conditions of thermal equilibrium, so establishing that the fluctuation is not caused by the current flow.

In our case the variable $u$ can be intended as the resistance $R$. In fact, M. Sourdin hypothesised that the fluctuation of the thermoionic current is due to fluctuations of resistance stemming from fluctuations of a number of free electrons on the metal [3].

These fluctuations are interpreted as the collective action of various microscopic RTN processes such as trapping and detrapping of conduction electrons.

Therefore the equation (12) can become:

$$S_R(f) \ \mu \ \mathbf{s}_R^2/f \qquad . \qquad (19)$$

It can be demonstrated using a simple model [7] that:

$$\mathbf{s}_R^2 \propto \frac{\langle R \rangle^2}{N} \qquad , \qquad (20)$$

where $N$ is the number of charge carriers in the sample.

If a steady current $I$ flows through the resistance $R$, we can multiply per $I^2$ the first and the second term of relation (19). Therefore, by also using the equation (20), we have:

$$S_V(f) \ \mu \ \frac{I^2 \mathbf{s}_R^2}{f} \propto \frac{\langle V \rangle^2}{N \ f} \qquad . \qquad (21)$$

The last term of equation (21) is just the *Hooge empirical formula* [8].

These results confirm that the equation (12) and the hypothesis of the resistance fluctuations are functional to deduce the relations (21).

**6.** $1/f^{g}$ *noise*

If pure *1/f* noise is generated by a Gaussian process then the form $1/f^{g}$ can be ascribed to a quasi-Gaussian behaviour. In other words, our process can be better described at low frequencies by a modification of the Gaussian expression (5) as:

$$P(\Delta u) \propto e^{-\frac{\Delta u^2}{2\mathbf{s}_{\Delta u}^2 g}} \qquad , \qquad (22)$$

where $\mathbf{g}$ is intended as a correction parameter close to one.

Therefore, the equation (9) becomes:

$$\Delta u^2 = 2\mathbf{s}_{\Delta u}^2 \ln \frac{f_{\max}^{g}}{f^{g}} \qquad . \qquad (23)$$

If we now introduce a new variable $\mathbf{n} = f^{g}$ the problem can be treated again in a Gaussian space so the equation (10) becomes:

$$\int_{\mathbf{n}}^{\mathbf{n}_{max}} S_u(\mathbf{n})\, d\mathbf{n} = \Delta u^2 \qquad\qquad (24)$$

and the equation (12)

$$S_u(f) \propto \mathbf{s}_{\Delta u}^2 / f^{\,g} \quad . \qquad\qquad (25)$$

## 7. 1/f noise on 1/f noise

Experimental results due to C.E.Parman et al. [9] and P.J. Restle et al. [12] reveal "second spectra" on the *1/f* noise power spectrum, which themselves have *1/f* behaviour.

It is possible to explain this result if we accept that these fluctuations are due to the Lorentzians with different relaxation times which generate the *1/f* power spectrum for short times. In fact, the power spectrum $S_2(\mathbf{w})$ of these fluctutions will be:

$$S_2(\mathbf{w}) = \frac{k}{\mathbf{w}} + \sum_t k_t \frac{\mathbf{t}}{1 + \mathbf{w}^2 \mathbf{t}^{\,2}} \quad , \qquad\qquad (26)$$

where $k$ and $k_t$ are a suitable constant of proportionality and the summation $\mathbf{S}_t$ is intended on all the Lorentzians which generate the *1/f* power spectrum.

Looking to equation (26), if $\mathbf{w} >> 1$ then $S_2(\mathbf{w})$ ? 0, and when $\mathbf{w} << 1$ then $S_2(\mathbf{w})$ ? $k/\mathbf{w}$ + const as it is expected .

## 8. White noise limit at low frequency

$S_u(\mathbf{w})$ exhibits a white spectrum when the number of the primary RTN processes is limited to such an extent as to enable the reaching of the inferior limit of frequency for the *1/f* noise within the time scale of the measurements.

In fact, if $\mathbf{t}_{max}$ is the maximum relaxation time of the Lorentzians generated by the primary RTN processes then for $\mathbf{w} << 1/\mathbf{t}_{max}$ $S_u(\mathbf{w})$ ® const [1][3]. This result is clearly confirmed when we look to the Lorentzians with the relaxation time less than $\mathbf{t}_{max}$ and, consequently, also to their summation.

Experimental evidence of this behaviour is reported in reference [13] when the flow of tiny grains in a hourglass is analysed, or in reference [4] where the flattening at the low frequencies of power spectra relating to *1/f* noise sources appears.

## 9. Gaussian noise without 1/f noise

If the generating processes are not RTNs but operate with the same probability distribution, the resulting distribution of the $\Delta u$ is once again Gaussian (*central limit theorem*). Thanks to the *law of large numbers*, even in this case the fluctuation $\Delta u$ will converge in an average process of frequency $f^*$.

This frequency $f^*$ cannot be defined as in equation (8) because now a relaxation time does not exist. Instead, we can introduce a frequency of $\Delta u$ if we look to the time interval $\Delta t$ adopted to perform the measurements, that is

$$f^* = \frac{P(\Delta u)}{\Delta t} \qquad \text{and} \qquad f^*_{max} = \frac{P(\Delta u)_{max}}{\Delta t} \qquad (27)$$

where $P(\Delta u)$ and $P(\Delta u)_{max}$ have the known meaning.

Each power $\Delta u^2$ is an average power in the time interval $\Delta t$ and relates to a process of frequency $f^*$. This power $\Delta u^2$ will have spectral components with a frequency greater than $f^*$ thanks to the Fourier theorem but the maximum frequency of these components cannot be $f^*_{max}$ of relationship (27). In fact, $f^*_{max}$ depends on $\Delta t$ while the measurement of $\Delta u^2$ in the frequency-domain involves the spectral components determined by the co-operation of the generating processes which have their own frequencies independently of $\Delta t$.

Therefore, in general the power spectrum $S_{u,\Delta t}(f)$ which generates $\Delta u^2$ must be integrated from $f^*$ to a frequency $f_{max}$ different from $f^*_{max}$. In practice, $f_{max}$ corresponds to the bandwidth of the generating processes. Therefore

$$\int_{f^*}^{f_{max}} S_{u,\Delta t}(f)\,df = \Delta u^2 = 2\sigma_{\Delta u}^2 \ln \frac{f^*_{max}}{f^*} \quad . \qquad (28)$$

The consequence of relationship (28) is that the equation (11) cannot be used, so pure *1/f* noise is not generated.

On the other hand, the structure composed of Lorentzians of the same amplitude and with a distribution of relaxation times which produce *1/f* noise cannot appear for short times if the generating processes are not RTNs. So, also from this point of view, the *1/f* noise is precluded

Therefore, a Gaussian noise can exist without necessarily being a *1/f* noise (e.g. white noise).

Concerning the filtering effects we have similar result to equation (16)

$$S_{u,\Delta t}(f) = S_u(f) \frac{\sin^2(\pi f \Delta t)}{(\pi f \Delta t)^2} \quad , \qquad (29)$$

where $S_u(f)$ is the limit of $S_{u,\Delta t}(f)$ when $f \cdot \Delta t \to 0$.

## 10. Serial 1/f noise

 A single two-state process is very common in nature, for example the sequential flow along the same path of identical objects such as electrons, photons, cars, etc.

 If $\boldsymbol{D}t$ is the minimum time necessary to detect one object which flows through a cross-section of the path, $N = 1/\boldsymbol{D}t$ represents the average number of attempts to find the object in the time unit.

 If $p$ is the probability of detecting one of these objects in time $\boldsymbol{D}t$ then the probability $P(n)$ to detect $n$ objects in unit time will follow the binomial distribution.

 When $N$ is large and the probability $p$ is not very small, $P(n)$ will become a Gaussian distribution, so

$$P(n) = \frac{1}{\boldsymbol{s}_n \sqrt{2\boldsymbol{p}}} \exp\left[-\frac{\left(n - \langle n \rangle\right)^2}{2\boldsymbol{s}_n^{\,2}}\right] \quad , \tag{30}$$

where $\langle n \rangle$ denotes the mean value of $n$ and $\boldsymbol{s}_n$ the standard deviation of the distribution.

 Solving equation (30) in the unknown $\boldsymbol{D}n^2 = (n - <n>)^2$, we obtain

$$\Delta n^2 = 2\boldsymbol{s}_{\Delta n}^{\,2} \ln \frac{P(\Delta n)_{\max}}{P(\Delta n)} \quad , \tag{31}$$

where $P(\boldsymbol{D}n) = P(n)$, $\boldsymbol{s}_{\boldsymbol{D}n} = \boldsymbol{s}_n$ and $P(\boldsymbol{D}n)_{max} = \dfrac{1}{\boldsymbol{s}_{\Delta n} \sqrt{2\boldsymbol{p}}}$.

 Thanks to the *law of large numbers*, the mean frequency $f$ of the fluctuations $\Delta n$ is proportional to their probabilities and therefore we can substitute $P(n)$ with $f$. In fact $f = P(\boldsymbol{D}n) / \boldsymbol{D}t$ and $f_{max} = P(\boldsymbol{D}n)_{max}/\boldsymbol{D}t$.

 $\boldsymbol{D}n^2$ represents the maximum power of the process of frequency $f$ whose spectral components have frequencies at the range of $f, f_{max}$.
Thus, introducing the power spectrum $S_n(f)$ we can write:

$$\int_f^{f_{\max}} S_n(f)\, df = \Delta n^2 \quad . \tag{32}$$

When $S_n(f) = 2\boldsymbol{s}_{\boldsymbol{D}n}^2/f$ , we have

$$\int_f^{f_{\max}} \frac{2\boldsymbol{s}_{\Delta n}^{\,2}}{f}\, df = 2\boldsymbol{s}_n^{\,2} \ln \frac{f_{\max}}{f} = \boldsymbol{d}n^2 \quad , \tag{33}$$

so

$$S_n(f) \boldsymbol{m}\ \boldsymbol{s}_{\boldsymbol{D}n}^2/f \quad . \tag{34}$$

## 11. Correlations of the 1/f noise

The mean  $<c>$  of the number of coincidences of $N$ similar RTN  processes which co-operate to produce 1/f noise, or the mean $<n>$  of the number of objects which flow sequentially in the unity of time, when N is the frequency of the attempts, is:

$$<c> = Np \quad \text{or} \quad <n> = Np , \qquad (35)$$

where $p$ is the occurrence probability of the events.
If $q$ represents the probability of not finding the event in an attempt :

$$\boldsymbol{s_{Dc}}^2 \, \boldsymbol{\mu} \, \boldsymbol{s_{Dc}}^2 = \boldsymbol{s_c}^2 = \; <c> q \quad \text{and} \quad \boldsymbol{s_{Dn}}^2 = \boldsymbol{s_n}^2 = \; <n> q \; . \qquad (36)$$

Thus, thank to equations (12) (34) and (36) :

$$S_u(f) \, \boldsymbol{\mu} \, <c> q \, / f = Npq/f \quad \text{and} \quad S_n(f) \, \boldsymbol{\mu} \, <n> q \, /f = Npq/f. \qquad (37)$$

This result tells us that the power spectrum of the 1/f noise can fluctuate in time if $<c>$ and $<n>$ fluctuate, too.  Fluctuations of $<c>$ can be originated by fluctuations of $N$ and/or $p$ while in the case of the sequential process only fluctuations of $p$ can determine fluctuations of $S_n(f)$, $N$  being steady "a priori".
 On the basis of these considerations the fluctuations of $S_u(f)$ or $S_n(f)$ in a given octave (or decade) are necessarily correlated to the fluctuations of the *1/f* noise power spectrum in other octaves (or decades).
  Experiments, performed on n-type doped hydrogenated amorphous silicon (a-Si:H) samples confirm that slow variations of the  *1/f* noise are strongly correlated over a broad range of frequencies [9]. In fact, these experimental results show that strong correlations of the average noise power as a function of time appear between the differing octaves.
  The existence of co-operative dynamics among the spectral components of noise power has also been observed in mesoscopically small CuMn spin glasses [10] and in simulations of kinetic Ising models [11].

## 12.  Conclusions

In this paper it has been demonstrated that  *1/f* noise appears when a co-operation exists among many and similar RTN processes or when a single binary process (e.g. the sequential flow of identical objects) operates with the same statistics.
  In fact,  thanks to the *law of the large numbers* applied to the Gaussian distribution of the resulting variable, this co-operative process can be conceived on average as a summation of spectral components which generate a *1/f* power spectrum.
 Moreover, secondary RTN processes can be viewed in the process for short times with such a distribution of relaxation times that their summation produces  *1/f* noise again.

On the other hand, Gaussian noise can exist without necessarily being *1/f* noise if the similar originating processes are not RTNs.

Various experimental results found in the literature are compatible with this model.

At least the so-called paradox of the *1/f* noise (the spectral density increases with decreasing frequency *f* as far down as it is possible to conduct measurements) arrives at an easy explanation. In fact, if we suppose, as an example, one hundred two-state fluctuators each having equal appearance probability in the two states and a relaxation time of 1 μs, the probability of a coincidence (during one microsecond) of all hundred RTN processes corresponds on average to one possibility in about every $2 \cdot 10^9$ years (the retained age of our Universe).

## References

[1] Sh. Kogan, *Electronic noise and fluctuations in solids*, Cambridge University Press, 1996.

[2] B.B. Mandelbrot, *Multifractals and 1/f noise*, Springer, 1999.

[3] M. Sourdin, J. Phys. Radium, 10 (1939) 188-189.

[4] R.F. Voss, Phys. Rev. Lett. A, 40 (1978) 913.

[5] R.F. Voss, J. Clarke, Phys. Rev. B, 13 (1976) 556.

[6] R.F. Voss, J. Clarke, Phys. Rev. Lett. , 36 (1976) 42.

[7] A. Ambrózy, *Electronic noise*, Mc Graw-Hill, 1982.

[8] F.N. Hooge, A.M.H. Hoppenbrouwers, Physica 45 (1969) 386.

[9] C.E. Parman, N.E. Israeloff, J. Kakalios, Phys. Rev. Lett. 69 (1992) 1097.

[10] N.E. Israeloff, G.B. Alers, M.B. Weissmann, Phys. Rev. B, 44 (1991) 12613.

[11] G.B. Alers et al., Phys. Rev. B, 36 (1987) 8429.

[12] P.J. Restle et al., Phys. Rev. B, 34 (1986) 4419.

[13] K.L. Schick, A.A. Verveen, Nature 251 (1974) 599.